\documentstyle[draft,psfig]{mn}
\newcommand{\mnref}[1]{\hangindent=0.5in \hangafter=1 #1 \par}

\newcommand{\mn}{MNRAS}

\newcommand{\aj}{AJ}
\newcommand{\apj}{ApJ}

\newcommand{\aaa}{A\&A}

\newcommand{\Lsolar}{\mbox{\,$ L_{\odot}$}}        
\def\gs{\mathrel{\raise1.16pt\hbox{$>$}\kern-7.0pt
\lower3.06pt\hbox{{$\scriptstyle \sim$}}}}
\def\ls{\mathrel{\raise1.16pt\hbox{$<$}\kern-7.0pt
\lower3.06pt\hbox{{$\scriptstyle \sim$}}}}

\title[Infrared Imaging Polarimetry of NGC1068]
{Near and Mid Infrared Imaging Polarimetry of NGC1068}
\author[S.L. Lumsden et al.]
{S.L. Lumsden$^{1}$, T.J.T. Moore$^{2}$, C. Smith$^{3}$, T. Fujiyoshi$^{3}$, 
J. Bland-Hawthorn$^{1}$ \and and M.J. Ward$^{4}$\\
{}$^1$ {\em Anglo-Australian Observatory, PO Box 296, Epping, NSW 2121,
Australia} \\
\hspace{1cm}{ Email -- sll@aaoepp.aao.gov.au, jbh@aaoepp.aao.gov.au}\\
{}$^2$ {\em Astrophysics Research Institute, 
Liverpool John Moores University, Byrom Street, Liverpool L3 3AF} \\
\hspace{1cm} {Email -- tjtm@staru1.livjm.ac.uk}\\
{}$^3$ {\em School of Physics, University College UNSW, Australian Defence
Force Academy, Canberra 2600, Australia}
 \\
\hspace{1cm} {Email -- craig@phadfa.ph.adfa.oz.au, txf@phadfa.ph.adfa.oz.au}\\
{}$^4$ {\em University of Leicester, Department of Physics and Astronomy}
 \\
\hspace{1cm}{ Email -- mjw@star.le.ac.uk}
}

\begin{document}
\label{firstpage}

\maketitle

\begin{abstract}
We present the results of a series of observations of the near- and mid-infrared
polarisation properties of the Seyfert 2 galaxy NGC1068.  Our data agree well
with previously published results in showing the need for a separate
polarisation mechanism in the near infrared apart from scattering.  We find
that the predictions of a simple model in which this component arises through
absorptive dichroism due to aligned grains within the extended warm
($\sim400$K) dust fits the data reasonably if the obscured background source is
itself due to dust emission (at temperature $>1000$K).  By considering the
change of polarisation with wavelength we show that the extinction to this hot
dust region is in the range $A_V=20-40$.  Consideration of the observed data
then leads us to the conclusion that if viewed face-on, NGC1068 would have a
strong near-infrared excess similar to Seyfert 1 galaxies.  Comparison with
other independent measures of the extinction to the active nucleus itself lead
us to the conclusion that the hot dust must provide screening equivalent to at
least $A_V=40$, and possibly much higher.  We speculate that this component
alone may be the `classical' torus discussed in terms of the unified model, and
the more extensive mid-infrared emission may arise from circumnuclear molecular
cloud material, and dust in the ionisation cones.

\end{abstract}


\section{Introduction}
The unified model for Seyfert galaxies proposes that all types of Seyfert
galaxy are basically the same, but that a dusty molecular `torus' obscures the
broad-line region (BLR) in many systems. The classification into Seyfert 1 or 2
(Seyfert 1: broad permitted lines and Seyfert 2: narrow permitted lines) then
depends on the inclination angle and hence on whether we can see the BLR.
Although the concept of the unified model existed before the discovery by
Antonucci \& Miller (1985) of broad H$\alpha$ emission in spectropolarimetry of
the nearby luminous Seyfert 2 galaxy NGC1068, that result was seen by many as
the major confirmation of the likely validity of the model.  Although other
Seyfert 2's have also been found to have broad permitted lines in scattered
light, this is by no means universal.  Indeed Heisler, Lumsden \& Bailey (1997)
found a strong correlation between the `warmth' of the IRAS 60$\mu$m to
25$\mu$m ratio and the ability to detect such scattered broad-line radiation.
Further studies that enhance our understanding of why some Seyfert 2s appear
consistent with the unified model and others do not are clearly therefore of
value.

In this context, infrared polarimetry of AGN provides valuable extra
constraints on the polarisation mechanisms active in Seyfert 2's.  Since the
unified model infers the presence of a dusty torus obscuring the AGN core, we
might expect that longer wavelength polarimetry will be able to probe deeper
into regions within the plane of the torus.  Hence infrared polarimetry may let
us see scattering in Seyfert 2's which is otherwise shielded from view.  In
addition, many Seyfert 2's with generally `warm' IRAS colours are known to have
polarisation rising into the near infrared (Young et al.\ 1996a).  The
mechanism generally invoked to explain such behaviour is linear absorptive
dichroism, in which aligned dust grains preferentially absorb one plane of
polarisation.  When such dust screens a bright background source, the
transmitted light is therefore polarised.  As shown by Hildebrand (1988),
observations of this aligned dust component at wavelengths where it emits as
well as at wavelengths where it absorbs can be used to study the nature and
geometry of the dust that so heavily obscures the AGN along our line of sight.

Since it is already known from the spectropolarimetric observations that
NGC1068 contains a hidden type 1 core, has relatively high polarisation for a
Seyfert 2 (Heisler, Lumsden \& Bailey 1997), and has rising infrared
polarisation (Young et al.\ 1995), it makes a perfect target in which to test
these ideas.  Its proximity (14.4\,Mpc; Tully 1988) and luminosity
also make it an ideal candidate.

There has already been considerable previous work on infrared polarisation in
NGC1068.  The most extensive early broad-band polarimetry data are reported by
Bailey et al.\ (1988), who found a strong rising red continuum in the 
near-infrared polarised flux and evidence for a gradual change in the
position angle of polarisation between the optical and near infrared.  They
explained these as a combination of polarisation by scattering and due to 
transmission of a strongly reddened background source through aligned dust 
grains.  They also found a much more marked change in polarisation angle
($\sim70^\circ$) between 4 and 5$\mu$m due to the same aligned grains being
seen in emission at longer wavelengths.  This large swing in polarisation 
is seen again in the spectropolarimetry of Young et al.\ (1995) who also 
included polarisation by aligned grains in their model, having found that
scattering fitted the data well below 1$\mu$m but was deficient 
beyond that wavelength in explaining both the polarised flux and the 
position-angle data.

In addition, Aitken et al.\ (1984) presented 10$\mu$m spectropolarimetry,
showing that the percentage polarisation remained constant through the
9.7$\mu$m silicate absorption feature.  This showed that the cause of the
10$\mu$m polarisation could not be absorption by aligned dust grains,
since that would have resulted in a rise in polarisation through the
absorption feature.  However, they were unable to decide between the other
alternatives: emission from aligned, optically thick (since the silicate
absorption feature seen is relatively weak) dust grains or polarised emission
from the active core itself.  The position angle of polarisation seen at 
10$\mu$m is consistent with that found by Bailey et al.\ (1988) at 5$\mu$m.

A consistent picture has begun to emerge from all these results.  The optical 
data near the nucleus can be explained by a combination of a dominant 
electron-scattering 
component and a lesser dust-scattering one.  The near-infrared polarisation
is a combination of electron scattering and the dichroic component modelled by
Young et al.\ (1995).  The longer wavelength data is then consistent with
emission from aligned grains, since the emission and absorption polarisation
position angles from the same grains are roughly orthogonal.

However, two aspects were clearly lacking in all of this work.  Since much of
the activity in NGC1068 occurs within 1--2 arcseconds of the nucleus, 
high spatial resolution observations are required to separate the different
polarisation mechanisms that may be present.  Some high-resolution imaging
polarimetry has been published by Young et al.\ (1996b) but this was obtained
in moderate seeing conditions.  The only other imaging data that exists in the
near infrared is that of Packham et al.\ (1997), but the resolution of this
data is $\ge1.2$ arcseconds.  All other previously published data were acquired
using aperture bolometer devices.  The second major lack is imaging polarimetry
in the mid-infrared.  Without such data it is not possible to relate the
observed near-infrared data to the behaviour at longer wavelengths with a high
degree of confidence.  We have therefore gathered high-resolution ($\sim 0.5$
arcsecond), complete near- and mid-infrared imaging polarimetry of the central
regions of NGC1068 using the AAT.  In section 2 we give details of these
observations, in section 3 we present the results of our observations, in
section 4 we analyse these results in terms of possible polarisation
mechanisms, in section 5 we discuss the dust distribution around the nucleus of
NGC1068 and finally in section 6 we present our conclusions.

\section{Observations}
All of the data presented in this paper were obtained on the nights of
9, 10 and 11 August 1995  and 11 October 1997
at the Anglo Australian Telescope.  The 10$\mu$m
data were acquired with the ADFA mid-infrared camera, NIMPOL
(Smith, Aitken and Moore, 1994), which
at the time used a 128$^2$ SiGa array, with a pixel scale of 0.25$''$/pixel.  
A cold wire grid and a warm rotating CdS half-wave plate were used as the
analyser for the polarimetry.  We used a standard broad band 8--13$\mu$m
filter for the observations, and chopped and nodded $\sim20''$ to sky
so that the object was always present on the array.  Extended emission
at 10$\mu$m is sufficiently weak (eg Telesco \& Decher 1988) that
this poses no problems for the nuclear polarimetry.  Approximately
7 hours of actual on-target data were acquired.  The image is diffraction
limited, and all data taken during non-photometric conditions have been
discarded.  We observed NGC1068 at
airmasses ranging between 1.2 and 1.8.

The near-infrared data were obtained on the nights of 11 August 1995 and 11
October 1997 with the common-user camera IRIS, which uses a 128$^2$ HgCdTe
array.  A warm rotating half-wave plate and a cold beam-splitting Wollaston
prism were used to obtain the linear polarimetry: the Wollaston allows for very
efficient polarimetric observations, and also allows reliable data to be taken
even in non-photometric conditions.  Again the pixel scale was 0.25$''$/pixel.
The seeing was estimated at 0.6$''$ during these observations.  After each
cycle of four waveplate positions we nodded the telescope to a separate sky
position since the galaxy is bright in the near infrared beyond the nucleus.
Approximately 20 minutes of data were obtained on source at each of J, H and
K$_n$.  Conditions were photometric for the H and J band observations, and
partly photometric for the K$_n$ observation.  The photometry given in Section
2 is derived solely from those frames taken when conditions were photometric.
Unfortunately, the conditions meant that it was difficult to obtain good `sky'
frames to subtract from the data at K$_n$ due to clouds reflecting thermal
emission from the ground.  The result is a higher than expected background
noise level in the polarisation images.  We therefore also observed NGC1068 at
H and K$_n$ on the night of the 11th October 1997 with the same setup to check
our results.  We obtained 16 minutes of on-source data at K$_n$ and 8 minutes
at H on this night.  The seeing was good for the K$_n$ data ($\sim$0.75
arcseconds), but poorer for the H band data ($\sim$1 arcsecond).  Conditions
were photometric during this night.  This is evident in a considerable decrease
in the noise in the polarised flux images.  The polarimetry derived from both
dates is consistent within the errors.

Flux calibration for the mid-infrared data was obtained from matching
observations of the mid-infrared standards BS6832 ($\eta$ Sgr; S(10$\mu$m) =
197 Jy), BS8636 ($\beta$ Gru; S(10$\mu$m) = 933 Jy) and BS7525 (S(10$\mu$m) =
93 Jy), and has an estimated uncertainty of 10\%.  Flux calibration for the
near-infrared 1995 data was derived from SA241--251 and for the 1997 data from
HD24849 (Carter \& Meadows 1995).


The polarimeters used with both IRIS and NIMPOL are very stable, and
instrumental effects result in errors as low as 0.1\% in polarisation and less
than 1$^\circ$ in position angle.  In order to check the performance of the
system with IRIS we also observed standard stars of known polarisation, as well
unpolarised standards and stars observed with a wire grid (of known
polarisation properties) in the beam.  This allows the zero-point of the
position angle to be set.  The resulting errors found were in line with the
expected performance.  For NIMPOL, the polarisation standard was the BN object
in Orion, assumed to have a polarisation angle of 118$^\circ$ at 10$\mu$m.

The near-infrared data were reduced as follows.  The data were flat-fielded
using dome flats.  Offset sky frames (several arcminutes from the source) were
obtained using the same setup as for the object data.  The frames for each
half-wave plate position were median-filtered to remove background sources.
The resultant four sky frames (one for each half-wave plate position) were
scaled to the median level of the sky within each group of four individual
object frames, and the result subtracted from the object frames (this technique
proved resilient to residual sky variations from tests carried out by
subtracting sky from individual sky frames themselves, which could otherwise
provide an unwanted additional background `noise' signal in the final
polarisation).  The resultant images were then registered and combined into
separate mosaics for each half-wave plate position.  These final mosaics were
combined to form Q and U Stokes images, and hence polarisation maps.  This
procedure is valid because of the dual-beam nature of the instrument.
Variations in sky brightness, or transparency, are correctly allowed for
because the frames are scaled when combined to form the Stokes parameters
ensuring that the total flux is conserved in any image.

The raw 10$\mu$m data is stored with the chop already subtracted but the
corresponding nod position is removed later in software.  The result is then
properly sky subtracted.  Flat fields were made from pairs of sky observations, 
one at high ($\sim1.6$) and one at low ($\sim1.0$) airmass, which were 
subtracted and normalised.  The sky-subtracted, flat-fielded data were 
registered to remove any small shifts in the object position with time, 
since the telescope was tracking without guiding.  The shift observed is 
only $\sim1$ arcsecond per hour.  Frames in which the background increased 
by more than 50\% compared to others in a given
sequence were discarded.  Such frames were generally affected by cloud.  The
final data were then grouped into Stokes Q and U parameters.  Note that since
NIMPOL uses a wire grid analyser and not a Wollaston prism, residual variations
in atmospheric
transparency can affect the final results.  We tested for this by breaking the
data into sequential blocks and compared the results.  We found no evidence that
cloud affected any of the data used in the final images.

\section{Polarisation Maps and Photometry}
\subsection{The near-infrared data}
The near-infrared polarisation maps are shown in Figure 1 and the corresponding
polarised-flux images (being flux multiplied by degree of polarisation) are 
shown in Figure 2.  We have derived an error estimate for the polarised 
flux from the scatter in the counts well away from the nucleus,
and in Fig.~1 only plot polarisation points which lie more than
3$\sigma$ above the mean background level.  
The images shown are derived from the 1997 data for H and
K$_n$, and from the 1995 data for J.  It is worth noting that the centroids of
the direct flux and polarised flux are identical at all three wavebands to
within less than one tenth of a pixel (0.02 arcseconds).  

The measured polarisation is given in Table 1.  The errors are derived by
measuring the polarisation on every individual subset of four half-wave plate 
positions.
We note that the measurement in the small aperture is highly sensitive to
factors such as seeing, telescope drift etc, and therefore has a much higher
error than the other values quoted.  The results are in reasonable agreement
with both Bailey et al.\ (1988) and Packham et al.\ (1997).  We believe that
the differences found by Packham et al.\ between their data and that of Bailey
et al.\ may be largely due to factors such as seeing variations.  The core of
the emission from NGC1068 is clearly strongly polarised, but the profile of
this core is narrow, so the resultant measured polarisation in a small aperture
is strongly dependent on factors such as seeing.

The most immediately obvious features present in Figure 1 are the 
centro-symmetric vector patterns characteristic of scattering seen both 
northeast {\em and} southwest of the core.  This
provides a direct constraint on the geometry of the central regions of NGC1068.
Since two scattering cones are seen in the present data, but only one in
in the optical and UV, the counter cone must be hidden by an extended screen
at shorter wavelengths.  The only likely candidate for this screen is the 
disc of the galaxy.  This implies that the cones in NGC1068 are
definitely not aligned along the plane of the galaxy, in agreement with
previous evidence that the northeastern cone illuminates the near side of 
the disc whereas the southwestern cone illuminates the far side 
(cf Figures 1 and 2 of Bland-Hawthorn et al.\ 1997).  In particular, 
maps of the [OIII] emission 
show that the southwestern emission region 
vanishes behind the larger-scale molecular ring that surrounds the nucleus 
at a distance of about 1kpc, whereas the northeastern emission is clearly 
visible.  The same phenomenon can also
explain the asymmetry in the X-ray emission (Wilson et al.\ 1992).

The polarisation vectors are clearly evident spread in an arc between the
common axis of the large-scale ionisation cones and radio emission (eg.\ Wilson
\& Ulvestad 1987, Evans et al.\ 1991; position angle $\sim30^\circ$) and that
of the 
inner radio jet (eg.\ Ulvestad et al.\ 1987; position angle $\sim10-30^\circ$).
Whether or not this is one single continuous structure is not absolutely clear
from the present data, however, which hint at there being two preferred
directions in which the scatterers lie, coincident with the two axes of the
large- and small-scale radio emission.  Data from HST may be invaluable in
determining the true situation.

Our data are in good agreement with the images presented by Young et al.\
(1996b).  The `shadow' they noted in their H band data is also present in our
data.  We plot the image profile along the 45$^\circ$ direction (ie along the
long axis of the images in Fig.~1) in Figure 3 for J, H and K.  Note that our 
sign
convention is opposite to that of Young et al., since positive offsets here are
in the same sense as in Figure 1.  In Fig.~3 we have used the H band data 
from 1995,
when the seeing was better than for the 1997 H band data.  The K band data from
both 1995 and 1997 are consistent, but we show the former to ensure consistency
with the other plots.  The clear dip in the profile seen at J and H at an
offset of $\sim1$ arcsecond agrees well with the H band profile plotted by
Young et al.\ along a slightly different position angle.  They attributed this
to the possible shadowing of the scattering region by the torus.  This is
consistent with the presumed orientation of the ionisation cones if the torus
is also aligned in the same sense.

Photometry in the three wavebands from both the August 1995 and October
1997 data are given in Table 2.  Errors are estimates based on the scatter
between the frames obtained.  The larger error on the K$_n$ data from August
1995 reflects the fact that not all of the data were photometric.  For the 1997
data, we used all of the K$_n$ data but only the first two minutes of the H
band data since the seeing deteriorated after that point (clearly visible from
the change in the individual image profiles).  We have used only data taken in
seeing better than 1 arcsecond.  Because of the limited field of view, it is
difficult to judge if the sky has been fully corrected for in these images.
Light from NGC1068 dominates the images at all wavebands.  We have adopted an
average correction for the sky based on the detected counts 10 arcseconds from
the nucleus.  If this is an overcorrection, the tabulated fluxes will
increase by less than 5\%.

We can compare these results with those of Glass (1995), who published
photometry over a twenty-year timespan in a 12 arcsecond aperture.  There are
several preliminary corrections that need to be made, however.  
Glass used a standard single-element photometer so that only 12 arcsecond
aperture data is published.  Because of the mask used in the imaging
polarimetry our data is restricted to a 6 arcsecond aperture at most.  Since
NGC1068 varies slowly with time in the near infrared (Glass 1995), we cannot
compare our data directly with other existing photometry.  However, we can use
such photometry to derive an estimate of the flux within a smaller aperture
from the Glass data.  We compared the Glass data at Julian date 2446314
with that from Bailey et al.\ (1988) in a 6-arcsecond aperture at 
JD$\sim2446450$.   This is a sufficiently small separation in time that 
the variability should be negligible.  We ascribed the difference to emission
in the annular ring
between 3 and 6 arcseconds radius from the peak.  As noted by Glass, the source
that varies has the colours of hot dust: it is essentially the same as the
bright core seen within 2 arcseconds of the nucleus in our images.  It is 
likely, therefore, that the annular ring has little or no variability in flux, 
since it
lies outside the hot-dust emission region.  Assuming this is correct, we
can subtract the derived annular flux from later data from Glass (JD 2449615)
which is closer in time to our August 1995 data.  Using this approximation we
derive a 6 arcsecond aperture flux density from the Glass data of 761mJy for K,
363mJy for H and 198 mJy for J.  The errors on these numbers are entirely
limited by the systematics of the approximations made, but it is likely that
these values are reliable to within 10\%.  The results are in good
agreement with our own 1995 data.

We can also compare our photometry with that of Packham et al.\ (1997).  Their
H band result would agree with ours if we made no additional sky correction.
Their K band result is discrepant with our data even if we make this assumption
(their result being $\sim400$mJy brighter than the value given in Table 2).

Lastly, we note that the 1995 and 1997 data are in reasonable agreement with
each other, and provide no strong evidence for a large change in brightness
at these wavebands over this time period.  Glass (1997) has reported
that NGC1068 may now be declining in brightness at near-infrared wavelengths,
although the magnitude of the decrease is small and roughly consistent
with our own results.

\subsection{The mid-infrared data}
The 10$\mu$m image is shown in Figure 4.  The effective diffraction limit of
the AAT at 10$\mu$m is $\sim0.6''$, and the dust emission is clearly not a
point source.  The data also clearly show that the polarisation structure
arises from an extended region (all points with a signal-to-noise ratio 
greater than
5 in the polarisation are shown).  In addition, there is a trend for the
polarisation to increase away from the peak of the flux.  The position angle of
the polarisation is essentially constant across the region within the errors at
$55^\circ\pm10^\circ$.  The measured aperture polarisation is given in Table 1.

Aitken et al.\ (1984) obtained aperture spectropolarimetry at 10$\mu$m of the
nucleus of NGC1068, showing a relatively featureless polarised flux spectrum.
They concluded that the likeliest cause was emission from warm aligned dust
grains, but could not rule out emission from the active nucleus itself.
However, their data conclusively rule out an origin for the polarisation in
absorption by dust since this would give a peak in the percentage polarisation
through the silicate absorption feature at 9.7$\mu$m.  The result they derived
was $1.39\pm0.09$\% polarisation at a rather poorly defined position angle
(measured on two separate occasions at 44.1$\pm$3.2$^\circ$ and
59.4$\pm2.2^\circ$in a 4.2 and 5.6 arcsecond diameter aperture respectively).
Our observed position angle is consistent with the latter, though the
measured polarisation is slightly higher in the 4.5 arcsecond aperture.

Dust emission can be seen up to 5$''$ from the nucleus, with the extended
emission largely aligned with the observed ionisation cones (the inner high
contours are also aligned with this direction as noted previously by Braatz et
al.\ 1993).  The measured flux density within a central 6 arcsecond aperture is
$24.7\pm2.1$Jy, and within a 10 arcsecond aperture $26.0\pm2.2$Jy.  The 
surface brightness at the lowest contour in Fig.~4 is $\sim100$mJy/sq. 
arcsecond.

\section{Analysis}
\subsection{Polarisation mechanisms and the location of the AGN core}
Young et al.\ (1995) found the dominant contribution to the polarisation at
short wavelengths ($\lambda<1.5\mu$m) is from scattering (mostly from
electrons).  We can check this by examining the deviation from the expected
centro-symmetric scattering pattern.  As shown by Capetti et al.\ (1995), if
only scattering is present then the average of the differences between the
observed position angle in any pixel and the expected position angle (which is
just the normal to the vector joining that pixel and the central source from
which light is scattered into our line of sight), should be zero.  Of course,
this is a semi-iterative process since the exact location of the scattering
centre is {\em a priori} unknown.  However, it is trivial to iterate towards a
solution for both the source of the scattered light and the deviation from the
expected centro-symmetric pattern.

Figure 5 shows the observed departure from a pure-scattering vector pattern for
all three wavebands.  It is clear that the deviation in the expected vector
pattern reaches a maximum at position angle $\sim115^\circ$ (ie for vectors
with predicted position angle of 30$^\circ$) at all wavelengths.  The observed
pattern is consistent with that expected if polarised flux with a constant
position angle were superposed on the observed scattered flux (see Section
4.2).  Indeed this pattern is clear from Figure 1 as well.  Only by masking out
a region 2 arcseconds in radius around the central nucleus do we actually see
no deviations from the expected scattering pattern.  It is clear therefore that
another mechanism contributes to the observed polarisation {\em at all
wavebands}.

We masked out all the data within 2 arcseconds of the peak when deriving the
actual scattering centre because of this additional mechanism that contributes
near the nucleus, as well as those data with signal-to-noise ratio in the
polarised flux less than 3$\sigma$ above the mean background.  We found we
could not determine the location of the scattering centre reliably by
projecting orthogonal vectors back from the polarisation vectors to find a
common centre.  Residual background results in a `blurring' of the polarisation
position angles, so that some of the points of intersection lie well away from
the nucleus.  We therefore calcuated, for each pixel in turn, the difference
between the observed position angle and the position angle expected if that
pixel were the centre.  We then found the global minimum of this function, and
this position is the scattering centre.  This is far more stable to the effects
noted, but at the cost of giving a less precise solution than would be possible
with the other method given better data.  This technique gave offsets from the
respective flux centroids of $0.11\pm0.18$, $0.09\pm0.21$ and $0.34\pm0.39$
arcseconds at J, H and K$_n$ respectively, which are all formally consistent
with no offset.

We note that Marco, Alloin \& Beuzit (1997) found the centroid of the K band
flux tentatively aligned with the location of the UV scattering centre (Capetti
et al.\ 1995), and the radio sources S1 and S2 (Gallimore et al.\ 1996), from
astrometry obtained using adaptive optics and offsets between the peaks of the
optical and IR light.  Our results are therefore entirely consistent with their
analysis.  We note that their result depends on determining a difference
between the position of the peak of the I band light and the peak of the K band
light.  Although it might seem unusual that the peak of the J band light should
therefore be coincident with the peak of the K band light (since I is closer to
J in wavelength than J is to K), the reason is simple.  The J band centroid is
dominated by the same polarised core that is seen at H and K even though the
contrast with the electron scattering cones is much less.  The same feature is
unlikely to appear at I, as is clear from Section 4.2.  It is the centroid of
this feature that determines the location of the AGN core as we show below.

Lastly, we consider the factors that may affect our solution.  These are
systematic global position angle errors, random position angle errors and the
role of seeing.  Systematic global errors in the position angle calibration are
constrained to be $<1^\circ$ from observations of polarised standard stars,
which results in an entirely negligible change in the location of the
scattering centre.  We checked for effect of random position angle errors using
a Monte-Carlo analysis of our observed data, drawing a new set of position
angles from the set by randomly shifting the angle consistent with the observed
error in the position angle. This had a negligible effect on the location of
the centre.  The effect of seeing is to blur the scattering pattern, especially
near the bright nuclear continuum source.  We have attempted to quantify this
by simulating a simple ionisation cone irradiated by a centrally peaked
Gaussian source, with noise levels throughout consistent with our data.  This
pattern was then smoothed to mimic the observed seeing, and rederived the
polarisation from the model Stokes Q and U images.  The result shows that
seeing effects the observed pattern only in the inner 0.5 arcseconds around the
flux centroid.  Since we exclude this data anyway, seeing clearly has little
affect on our result either.  Therefore we can be confident that our result is
correct, and that the scattering centre is coincident with the K band
flux centroid.

\subsection{The nature of the polarised core}
As indicated by the analysis above, there is at least one other polarisation
mechanism present within $\sim1-2$ arcseconds of the peak of the flux.  Young
et al.\ (1995) originally proposed that this was absorptive dichroism on the
basis that no scattering law appeared to fit their spectropolarimetry.  That
suggestion was consistent with the results of Bailey et al.\ (1988) who also
concluded that the cause of the $\sim70^\circ$ change in position angle between
the near infrared and the mid infrared is the switch from emission to
absorption by aligned grains as wavelength decreases.  A lower limit to the
temperature of the dust providing the screen can be set by the observation of
Bailey et al.\ (1988) that the position angle of polarisation changes somewhere
between 4 and 5$\mu$m, from which we conclude that T$_{dust}\gs350$K.

The fact that the change in polarisation position angle is not closer to
90$^\circ$ may indicate one of two possibilities: first that the background
source seen at $\lambda < 4\mu$m contributes to some extent at 10$\mu$m too but
is polarised itself with a different position angle (so at short wavelengths,
the position angle is determined by the combined effects of dichroic absorption
and polarised emission from the background, and at 10$\mu$m from emission from
both); second, that there is yet cooler dust that is still shielding the
10$\mu$m emission, so that the observed position angle is a blend of the
emission and absorption processes (one test of this may be obtained by
determining the polarisation in the sub-mm regime).

All our data are consistent with the background source being hot dust with
temperatures near the sublimation point (see below), where grains are unlikely
to be aligned by a magnetic field.  Therefore {\em if} the grains are aligned
by a magnetic field, the first possibility is ruled out.  Alternative
mechanisms such as alignment due to radiation pressue in the ionisation cone
may give rise to a similar pattern (eg radiation pressure: Dopita et al.\ 1998)
but it is difficult to see how these can give rise to alignments in both hot
($>1000$K) and warm/cool ($\ls400$K) dust components.  It is also known that
NGC1068 has a prominent silicate absorption feature at 9.7$\mu$m (Roche et al.\
1984), indicating the last explanation may be the most probable.  If it is
true, we would predict that the observed position angle may change through the
silicate absorption feature from the $45^\circ$ measured at 5$\mu$m by Bailey
et al.\ to the $\sim55^\circ$ we measure.  Unfortunately the data of Aitken et
al.\ (1984) are not sufficiently accurate to measure such small shifts.

We have modelled the wavelength dependence of the observed polarised flux in
the core.  The model is simple and assumes only three parameters that are
allowed to vary.  These are the temperature of the dust, T$_{dust}$, the
emissivity of the dust (assumed to vary as a power law, $\lambda^{-n}$), and
the extinction to this dust source as measured in the K band, $A_K$.  The
polarised flux is assumed to be represented by an optically thin grey body with
this temperature and emissivity, scaled by an appropriate Serkowski
polarisation law (Serkowski, Mathewson \& Ford 1975 and references therein) 
and then reddened using a standard
$\lambda^{-1.75}$ extinction law.  
The Serkowski law has the form
\[ p(\lambda)/p_{max} =
\exp\left[-K\ln^2\left(\lambda_{max}/\lambda\right)\right] \] where we have
adopted `typical' values for moderately extinguished sources in our galaxy
(Whittet et al.\ 1992) of $K=1.15$ and $\lambda_{max}=0.6\mu$m (the effect of
changing $\lambda_{max}$ and $K$ is small for the wavelength range of
interest).

The results are indicative only given the few data points we are trying to
fit. It may be possible, given small aperture spectropolarimetry between 1 and
4$\mu$m, to provide better discrimination between the models.  However, a
typical `good fit' to the data can be obtained with the following parameters:
T$_{dust}=1200^{+300}_{-200}$, $n=1.5\pm0.5$ and $A_K=1.5\pm0.5$ (or
$A_V=17\pm6$).  The errors given here are not formal uncertainties but
represent the range of values that produce reasonable fits to the three data
points.  The results are also correlated in the sense that larger $n$ gives
lower values of T$_{dust}$, and larger values of $A_K$ require larger values of
T$_{dust}$.  However, it is clear that an extinguished hot dust source can
explain the observed data.  Since we might expect the hottest dust present to
be near the sublimation temperature, the `likeliest' parameters are those with
high T$_{dust}$ and $A_K$ (ie T$_{dust}$=1500K and $A_K=2$).

We also considered a simple power law for the underlying source rather than a
grey body.  A reasonable fit can again be derived with similar $A_K$ assuming
the underlying source varies as $\lambda$, which, however, is inconsistent with
typical synchrotron spectra.  For the dust temperature derived, there is
essentially no flux below 1$\mu$m, explaining why and J and K$_n$ centroids are
similar to each other, and not to I band data.

We can estimate what fraction of the total flux at J within a 3 arcsecond
aperture arises from the dust assuming the model with both large T$_{dust}$ and
$A_K$ is correct.  If we use the result of Thatte et al.\ (1997) that dust
contributes 90\% of the direct flux at K, and our model above, we predict that
dust will contribute $\sim50$\% of the total flux at H, and $\sim10$\% at J.
This is in agreement with our findings in Section 4.1 that even at J there is a
source of polarisation other than scattering, and that the J flux centroid is
coincident with the K flux centroid, since the centroids at both wavebands are
determined largely by the centrally peak dust component.  The same hot dust
contributes similar flux density at 2 and 10$\mu$m, or $\sim5$\% of the total
at 10$\mu$m.  Since in reality there will be a continuous range in dust
temperature with distance from the source, it is likely that up to 25\% of the
10$\mu$m emission can come from dust hotter than 1000K.

Lastly, we can obtain an estimate of the position angle and dimensions of the
inner regions of the obscuring material from our data.  The position angle of
the symmetry axis of this `torus' is $\sim30^\circ$ (90$^\circ$ from the
deviation in the scattering pattern shown in Figure 5).  This agrees well with
previous estimates of the position angle of the torus (Young et al.\ 1995,
1996b, Miller et al.\ 1991), and of the large-scale radio emission (Wilson \&
Ulvestad 1987).  The observed data are also marginally suggestive of an
extended source aligned perpendicular to this, since the profiles along this
direction and along the cone differ in extent by $0.1\pm0.05$ arcseconds.

To derive better estimates of the size of the resolved inner-torus region we
created a simple model in which a polarised core was superposed onto background
scattered emission.  The relative strengths of the core and background were
taken from the actual K$_n$ data.  The polarisation was also taken from the
observed data.  We then convolved this model with the observed seeing.  The
resultant polarisation maps, though highly idealised, could be compared with
the data to determine the actual extent of the polarised core.  The results
indicate that this region has approximate extent $7\times15$pc.  We show in the
next section that this extent is consistent with the observed mid-infrared
emission from the ionisation cones.  Clearly, high spatial resolution
($\sim0.1$arcseconds or better) IR imaging polarimetry that can sample the fine
structure in this region is desirable to estimate the true size of the torus.

\subsection{Mid-infrared emission}
As shown by both Braatz et al.\ (1993) and Cameron et al.\ (1993), it is
possible to relate the observed dust temperature and its distance from the
heating source to the source luminosity, after making suitable assumptions
about the grain properties and assuming that the dust sees the source
directly).  Cameron et al.\ derive the following equation, assuming an
emissivity proportional to $\lambda^{-1.5}$:

\[ \left(\frac{R}{0.14{\rm pc}}\right) =	
	\left(\frac{0.05\mu{\rm m}\times L}
	{a \times 1.5\times10^{11}\Lsolar}\right)^{0.5} 
	\left(\frac{T_{dust}}{1500{\rm K}}\right)^{-2.7}.
\]

Here $a$ is the grain size in microns, $L$ is (effectively) the luminosity of
the core and T$_{dust}$ the dust temperature.  Clearly, for fixed $R$ and $L$,
the observed dust temperature is weakly inversely proportional to the grain
size.  Therefore, we might expect that it is the physically smallest grains
(with radii$\sim0.005\mu$m) that give rise to the hottest dust emission 
at any point and we adopt this factor in the following analysis.

First we consider the dust in the ionisation cone.  Although we do not have an
estimate of the dust temperature in the ionisation cones, we can assume a lower
limit, which then gives a lower limit to the luminosity.  The fact that we see
dust emission at all at 10$\mu$m implies T$_{dust}$ $\gs 150$K, just from a
consideration of the sharpness of the drop in flux density with wavelength of a
suitable grey body.  If this dust is radiatively excited by the nucleus, this
implies that the dust in the ionisation cones sees a luminosity of at least
$\sim3\times10^{11}$\Lsolar.  This is a factor of two higher than the estimated
bolometric luminosity along our line of sight and may indicate anisotropic
emission.  This is not a new conclusion; Miller et al.\ (1991) found that the
properties of the electron scattering region were also consistent with
anisotropic emission from the core, as did Young et al.\ (1997).

Returning to the issue of the compact hot-dust emission discussed above, we see
that for the same parameters, dust can be heated to 600K at a distance from the
core of 8pc.  Note we assume a lower limit to the temperature for the
marginally `extended' emission at K, since the 1500K dust visible at all three
near infrared wavebands is likely to be an unresolved point source in our data
(there is no visible extent at H and J so we cannot apply this analysis).
Given the many assumptions that have gone into this simple analysis, the fact
that this distance agrees with the limits set by our data is highly
encouraging.  The more extended dust around the nucleus should be dealt with
using a proper radiative transfer model, since the source is likely to be
screened by the hotter dust in that case (cf Pier \& Krolik 1993, Efstathiou et
al.\ 1995).  It is interesting to note that the same parameters
predict a dust temperature of approximately 100K at the radius of the
large scale molecular ring seen in CO band images of NGC1068, which may 
just be warm enough to explain the excess 10$\mu$m emission seen in that
ring coincident with the ionisation cones (Bland-Hawthorn et al., 1997).

We modelled the observed 10-$\mu$m polarisation structure as a combination of a
point source (a reasonable approximation for the hot dust) and an extended
component (warm dust).  We assumed that the point source was completely
unpolarised.  The same results are obtained if the point source is polarised
along a different axis to the extended component however.  We convolved the
point source with a suitable Airy function and subtracted this profile from the
10-$\mu$m data until the percentage polarisation was constant across the
nucleus.  The best results are obtained if we allow the point source to have
peak height ~25\% of the observed peak flux.  The resulting polarisation
pattern after subtraction then has p$\sim1.4$\%.  This result is also
consistent with that expected from the model outlined in Section 4.2, where we
found that between 5 and 25\% of the emission at 10$\mu$m could arise in dust
at temperature greater than 1000K (ie dust where the alignment is likely to be
destroyed).

\section{Comparison of Mid and Near Infrared Data}
Young et al.\ (1995) modelled the polarised core of NGC1068 as arising due to
absorptive dichroism, with an extinction to the emitting source of A$_V\sim45$.
However, this was based on the assumption that the contribution of the dust to
the continuum emission seen in their $3''\times3''$ aperture is $\sim15$\% at
H, whereas Origlia et al.\ (1993) estimate this contribution is $\sim30$\% (in
a slightly larger aperture) and Thatte et al.\ (1997) derive an estimate of
$65\pm5$\% within a similar sized aperture.  At K, Thatte et al.\ find that
dust contributes $89\pm5$\% of the light within the same aperture, which is
also larger than the value given by Origlia et al.\ (70\%) within a slightly
larger aperture.  It is worth comparing these values with the crude estimates
derived in Section 4.2.  There we found $\sim50$\% of the total H band light
should be due to dust which is midway between the Thatte et al.\ and Origlia et
al.\ estimates, but given the crudity of our procedure probably consistent with
both.  Packham et al.\ (1997) also estimated the required $A_V$ to explain the
observed data using a similar method to that given below, though placing the
emphasis on the residual polarised flux after scattering had been allowed for,
and derived A$_V\sim35$.  It is useful to consider this problem independently
of the `best fit' scattering model to see what limits can be placed.

The shape of the near-infrared spectrum is very sensitive to the dust
temperature, assumed emissivity and extinction to the emission region as shown
in Section 4.2. As noted there, changing any of these parameters has an effect
on the others since they are correlated in any fit.  It is therefore very
helpful to derive independent estimates of any of these quantities.  One method
of deriving an independent estimate of the extinction is to compare the
relative polarisation at 10$\mu$m and K.  The dust that shields the hot
emission source evident in the K band is clearly cooler (and more spatially
extensive from our 10$\mu$m image) as noted above.  As shown by
Hildebrand (1988), the {\em intrinsic} percentage
polarisation seen due to aligned dust grains both in emission and 
absorption can be related according to
\[p_{em} = {p_{abs}}/{\tau_{\lambda(abs)}}, \]
if the result is taken at the same wavelength for both.  This is clearly
impractical, and a useful relation between absorptive polarisation at K and
emissive polarisation at 10$\mu$m must depend on the wavelength dependence 
of the dust cross-sections.

There are several other key assumptions in deriving a useful result from this
simple relationship.  First, it is assumed that scattering is not important at
short wavelengths.  This is true if we use a sufficiently small aperture near
the peak of the flux at K$_n$, and at H to a lesser extent.  It also assumes
that the 10$\mu$m emission is optically thin which is probably a reasonable
approximation (note the fact that there is an observed silicate absorption
feature as shown by Roche et al.\ (1984) does not violate this condition).  The
same grains must contribute at both wavelengths, and the wavelength dependence
of the absorption/emission cross-sections are taken from the observed
dependence of the known extinction law, from which $\tau_K\sim1.7\tau_{10}$,
and hence the same for the cross-sections.  Therefore, we apply a correction of
0.6 to the observed $K_n$ band polarisation.

We cannot use the observed values directly in this formula, since it relates
solely to the properties of the dust.  We can, however, correct these values,
by estimating what fraction of the polarised light is due to scattering rather
than dichroism, and what fraction of the direct light is due to stars rather
than dust emission.  For the former, we have derived aperture photometry on the
polarised flux images within an aperture of diameter 3 arcseconds.  We subtract
off the extended contribution underlying the central source, assuming it is due
to scattering (since this is where the scattered broad permitted lines arise).
The result is assumed to be the polarised flux due to dichroic absorption of a
background source.  Of course, if the `background correction' is not
representative of the scattering properties near the core (for example, if the
electron scatterers themselves are sharply peaked near the core) this will be
an overestimate of the polarised flux due to dichroism.  We can be confident
this is not the case, however, just from the observed behaviour of the position
angle of polarisation near the core.  This essentially shows that within the
innermost 1 arcsecond, scattering cannot dominate the light.  The derived
corrections are small for the region within 1.5 arcseconds of the core, and we
find that $90-95$\% of the polarised flux in this region arises in the strongly
peaked feature seen in Figure 3.

For the stellar contribution to the direct light, we use the values given by
Thatte et al.\ (1997) but note that there appears to be an inconsistency
between the results of Thatte et al.\ and Origlia et al.\ (1993).  The
difference in the result is small for the K-band data (A$_V$ is $\sim3$
magnitudes larger), but significant if we use the H-band data.  As a result we
do not consider the H-band data in detail, but note that a similar result (in
terms of A$_V$) to that derived at K is obtained if we use the lower Origlia et
al.\ value.

We also need to correct for possible extra sources that contribute to the
10$\mu$m polarisation.  As noted above it is possible that the hot dust
contributes 25\% of the light but none of the polarisation at 10$\mu$m, and we
have removed a point-source contribution at this level.  The main uncertainty
in the mid-infrared polarisation, however, is the dust opacity.  Assuming the
emitting regions are partially opaque implies that some of the emissive
polarisation is itself absorbed as noted before.  Hence the observed
polarisation is actually reduced since the net effect of the absorption is to
reduce the contrast between the axes of maximum and minimum polarisation (which
is also the likely reason why the position angles are not orthogonal as noted
in Section 4.2).  Therefore, the result derived here should be thought of as an
upper limit.  

Therefore, assuming that $p_{abs}\gs 1.4$\% and using $p_{em}\sim 4.7$\%, we
derive $\tau_K\ls3.3$, or $A_V\ls37$ assuming a standard IR extinction law.
This sets the upper limit to the values the data allow.  Since the estimated
extinction to the AGN itself is $>80$ (Jackson et al.\ 1993), the cool dust
is clearly only providing a fraction of the observed extinction.  Our result is
much larger than the observed optical depth in the silicate absorption feature
(Roche et al.\ 1984 derive $A_V\sim8$), but reasonably consistent with the C-H
band stretch absorption seen at 3.4$\mu$m (Bridger et al.\ 1994 find
$A_V\sim22$) the fits derived for the near-infrared polarised flux in Section
4.2 (where we found $A_V\sim23$ a reasonable fit) and the value derived by
Packham et al.\ (1997) ($A_V\sim35$).  The discrepancy with the silicate
absorption feature is most likely explained by infilling from the hot dust
emission washing out the absorption feature, as shown by full radiative
transfer models (Efstathiou et al.\ 1995, Pier and Krolik 1993).  Clearly
therefore, if the hot dust is cospatial with the AGN core, then that dust
itself must provide the bulk of the extinction to the core along our line of
sight.

\section{Conclusions}
We have obtained new near- and mid-infrared imaging polarimetry of NGC1068.
Our data are consistent with other observed results, but have generally higher
spatial resolution than anything yet published.  We can see clearly from our
data that scattering alone fails to fit the observed data at all wavebands, and
not just at K as noted previously by Young et al. (1995).  We have shown how
this deviation can be explained by absorptive dichroism of a background hot
dust source, and find that at least part of the hot dust must have a
temperature $>1000$K.  The screening dust must have temperature $>350$K, since
the observed change in position angle expected when moving from aligned grains
emitting to aligned grains absorbing arises at $\sim4-5\mu$m.  It is likely
that even cooler dust is also present which is shown only by the fact that the
change in position angle in the infrared is not exactly 90$^\circ$.

We have derived independent estimates of the extinction to the background hot
dust source seen through the obscuring screen.  The results, from fitting the
observed near-infrared polarisation data and from comparing the near- and
mid-infrared polarisation give results that are consistent with $A_V=20-40$.
This is considerably less than the estimate for the extinction through the
observed molecular material to the AGN core itself (eg Jackson et al.\ 1993
find that A$_V=80$ from consideration of the optical depth observed in the
circumnuclear molecular gas).  The large visual extinction to the hot dust
however explains why previous authors have noted the absence of a near-infrared
excess in NGC1068 as typically seen in Seyfert 1's (eg Edelson \& Malkan 1986,
Cameron et al.\ 1993).  If the hot dust were unobscured, the combined spectrum
of it and the more extensive warm dust would indeed peak near 5$\mu$m.  It may
be that the hot dust component is actually the `classical' torus, whereas the
warm dust is distributed more widely and diffusely throughout the circumnuclear
molecular clouds and the narrow line region.  This is in line with the results
of Heisler et al.\ (1997) who found a strong correlation between the size
of the electron scattering region and the obscuring material in a sample
of far-infrared selected Seyfert 2's.  

It is clear from our results that near-infrared polarimetry offers a
potentially valuable insight into the nature of the obscuring material.  In
particular, it allows us to determine the properties of the warm dust that is
nearest the AGN core.  The clearest need for the future is high spatial
resolution spectropolarimetry to define the polarisation behaviour as a
function of wavelength, and for polarisation predictions to be included in
radiative transfer modelling of the dust absorption/emission near AGN.  The
former will be especially useful when applied to techniques such as that
discussed in Section 4.2, as well as greater information of the overall
polarisation properties of dust absorption features, which may show weak
features in the change in position angle not previously considered.  There is
also clear scope for these methods to be applied to other nearby Seyfert 2s
with warm far-infrared colours.

\vspace*{3mm}

\parindent = 0pt

{\bf References}\par
\mnref{Aitken, D.K., Bailey, J.A., Briggs, G., Hough, J.H., Roche, P.F., 1984,
	Nature, 310, 660}
\mnref{Antonucci, R., Miller, J., 1985, \apj, 297, 621}
\mnref{Bailey, J.A., Axon, D.J., Hough, J.H., Ward, M.J., McLean, I.S., 
	Heathcote, S.R., 1998, \mn, 234, 899}
\mnref{Bland-Hawthorn, J., Lumsden, S.L., Voit, G.M., Cecil, G.N.,
	Weisheit, J.C., 1997, 
	In Gallimore, J., Tacconi, L., eds, 
	The Proceedings of the Workshop on NGC1068, 
	ApSS, 248, 177}
\mnref{Braatz, J.A., Wilson, A.S., Gezari, D.Y., Varosi, F., Beichman, C.A.,
	1993, ApJ, 409, L5}
\mnref{Bridger, A., Wright, G.S., Geballe, T.R., 1994, in McLean, I., ed.,
	Infrared Astronomy with Arrays, Kluwer, Dordrecht, p537}
\mnref{Cameron, M., Storey, J.W.V., Rotaciuc, V., Genzel, R., Verstraete, L.,
	Drapatz, S., Siebenmoregen, R., Lee, T.J., 1993, \apj, 419, 136}
\mnref{Capetti, A. Axon, D.J., Macchetto, F., Sparks, W.B., Boksenberg, A.,
	1995, \apj, 446, 155}
\mnref{Carter, B.S., Meadows, V.S., 1995, \mn, 276, 734}
\mnref{Dopita, M.A., Heisler, C.A., Lumsden, S.L., Bailey, J.A., 1998,
      \apj, 498, 570}
\mnref{Edelson, R.A., Malkan, M.A., 1986, \apj, 308, 59}
\mnref{Efstathiou, A., Hough, J.H., Young, S., 1995, \mn, 277, 1134}
\mnref{Evans, I.N., Ford, H.C., Kinney, A.L., Antonucci, R.R.J., Armus, L.,
 	Caganoff, S., 1991, \apj, 369, L27}
\mnref{Gallimore, J.F., Baum, S.A., O'Dea, C.P., 1996, \apj, 464, 198}
\mnref{Heisler, C.A., Lumsden, S.L., Bailey, J.A., 1997, Nature, 385, 700}
\mnref{Hildebrand, R.H., 1988, QJRAS, 29, 327}
\mnref{Jackson, J.M., Pagalione, T.A.D., Ishizuki, S., Nguyen,-Q-Rieu, 1993,
	\apj, 418, L13}
\mnref{Marco, O., Alloin, D., Beuzit, J.L., 1997, \aaa, 320, 399}
\mnref{Miller, J.S., Goodrich, R.W., Mathew, W.G., 1991, \apj, 378, 47}
\mnref{Origlia, L., Moorwood, A.F.M., Oliva, E., 1993, A\&A, 280, 536}
\mnref{Packham, C., Young, S., Hough, J., Axon, D., Bailey, J., 1997, MNRAS, 
	288, 375}
\mnref{Roche, P.F., Aitken, D.K., Phillips, M.M., Whitmore, B., 1984, \mn, 
	207, 35}
\mnref{Serkowski, K., Mathewson, D.S., Ford, V.L., 1975, \apj, 196, 261}
\mnref{Smith C.H., Aitken D.K., Moore T.J.T., 1994, in Crawford D.L., 
	Craine E.R., eds, Proc. SPIE 2198, Instrumentation in Astronomy VIII, 
	SPIE, Bellingham, 2198, p. 736}
\mnref{Telesco, C.M., Decher, R., 1988, \apj, 334, 573}
\mnref{Thatte, N., Quirrenbach, A., Genzel, R., Maiolino, R., Matthias, T., 
	 1997, \apj, 490, 238}
\mnref{Tully, R.B., 1988, Nearby Galaxies Catalog, Cambridge University Press}
\mnref{Ulvestad, J.S., Neff, S.G., Wilson, A.S., 1987, \aj, 93, 22}
\mnref{Whittet, D.C.B., Martin, P.G., Hough, J.H., Rouse, M.F., Bailey, J.A.,
	Axon, D.J., 1992, \apj, 386, 562}
\mnref{Wilson, A.S., Ulvestad, J.S., 1987, \apj, 319, 105}
\mnref{Wilson, A.S., Elvis, M., Lawrence, A., Bland-Hawthorn, J., 1992, \apj,
	391, L75}
\mnref{Young, S., Hough, J.H., Axon, D.J., Bailey, J.A., Ward, M.J., 1995,
       \mn, 272, 423}
\mnref{Young, S., Hough, J.H., Efstathiou, A., Wills, B.J., Bailey, J.A., Ward,
	M.J., Axon, D.J., 1996a, \mn, 281, 1206}
\mnref{Young, S., Packham, C., Hough, J.H., Efstathiou, A., 1996b, \mn, 283, L1}
\mnref{Young, S., Hough, J.H., Axon, D.J., Bailey, J.A., in 
	Wickramaisnghe, S., Ferrario, L., Bicknell, G., eds, 
	Accretion Phenomena and Related Outflows, ASP Conference Series
	 121, 280}
\newpage

\begin{tabular}{ccc}
\multicolumn{3}{c}{J band polarisation}\\
Aperture Diameter  & Polarisation  & Position Angle \\
(arcsec) & (percent) & (degrees) \\
2 & $2.25\pm0.25$ & $106.8\pm2.0$\\
4.5 & $1.59\pm0.15$ & $105.3\pm1.0$\\
6 & $1.35\pm0.10$ & $103.9\pm1.0$\\
\end{tabular}
\begin{tabular}{ccc}
\multicolumn{3}{c}{H band polarisation}\\
Aperture Diameter  & Polarisation  & Position Angle \\
(arcsec) & (percent) & (degrees) \\
2 & $3.87\pm0.20$ & $117.0\pm2.0$\\
4.5 & $3.06\pm0.10$ & $115.0\pm1.0$\\
6 & $2.58\pm0.05$ & $114.5\pm0.5$\\
\end{tabular}
\begin{tabular}{ccc}
\multicolumn{3}{c}{K$_n$ band polarisation}\\
Aperture Diameter  & Polarisation  & Position Angle \\
(arcsec) & (percent) & (degrees) \\
2 & $4.57\pm0.5$ & $120.2\pm2.0$\\
4.5 & $4.19\pm0.08$ & $120.3\pm0.5$\\
6 & $4.07\pm0.08$ & $119.6\pm0.5$\\
\end{tabular}
\begin{tabular}{ccc}
\multicolumn{3}{c}{N band polarisation}\\
Aperture Diameter  & Polarisation  & Position Angle \\
(arcsec) & (percent) & (degrees) \\
2 & $1.30\pm0.05$ & $49\pm3$\\
4.5 & $1.80\pm0.05$ & $57\pm6$\\
\end{tabular}

{\noindent \bf Table 1:} The measured polarisation within a circular
aperture around the peak of the observed total flux.

\begin{tabular}{cc}
Aperture Diameter  & Flux Density\\
(arcsec) & (mJy) \\
\multicolumn{2}{c}{J band photometry}\\
3  & 93$\pm$10 \\
6  & 198$\pm$10 \\
\multicolumn{2}{c}{Aug. 1995 H band photometry}\\
3 & 243$\pm$10 \\
6 & 390$\pm$15 \\
\multicolumn{2}{c}{Oct. 1997 H band photometry}\\
3 & 240$\pm$15 \\
6 & 420$\pm$15 \\
\multicolumn{2}{c}{Aug. 1995 K$_n$ band photometry}\\
3 & 650$\pm$30 \\
6 & 800$\pm$30 \\
\multicolumn{2}{c}{Oct. 1997 K$_n$ band photometry}\\
3 & 553$\pm$20 \\
6 & 785$\pm$10 \\
\end{tabular}

{\noindent \bf Table 2:} The measured photometry for NGC1068 in a circular
aperture around the peak of the flux.  


\onecolumn

\newpage

\parindent=0pt
{\bf Figures}

\begin{center}\begin{minipage}{5in}
\psfig{file=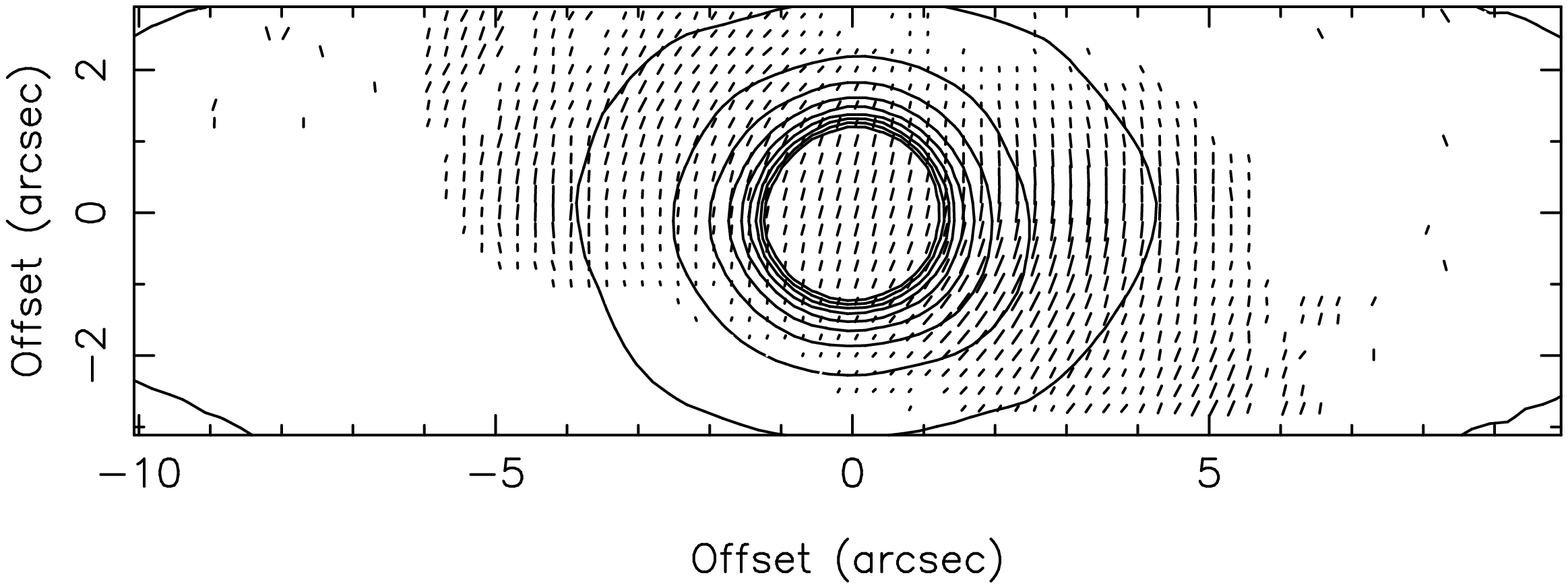,angle=-90,width=5in,bblly=20pt,bbury=720pt,bbllx=190pt,bburx=480pt}

\vspace*{-1.0in}
\hspace*{0.6in}\psfig{file=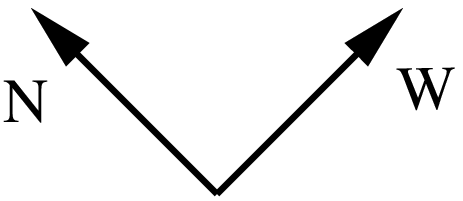,width=1.5cm}
\vspace*{0.8in}

\psfig{file=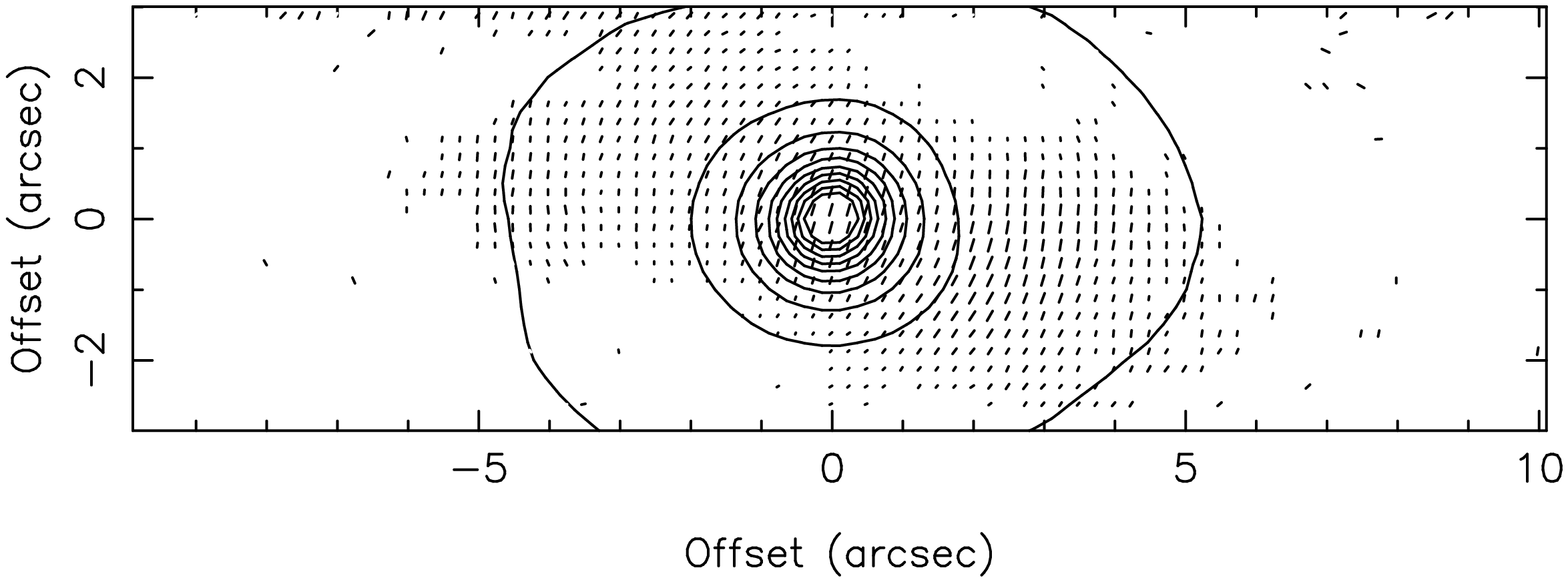,angle=-90,width=5in,bblly=20pt,bbury=720pt,bbllx=190pt,bburx=480pt}

\psfig{file=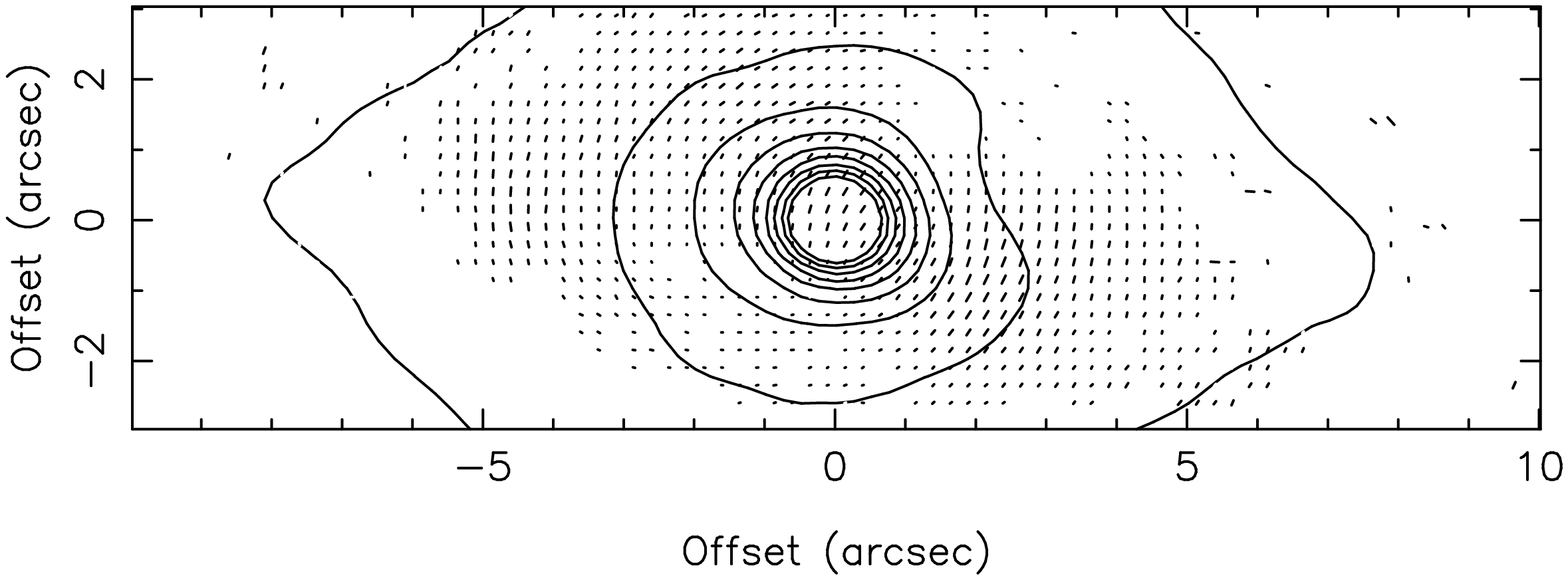,angle=-90,width=5in,bblly=20pt,bbury=720pt,bbllx=190pt,bburx=480pt}
\end{minipage}\end{center}
{\bf Figure 1:} Near infrared polarisation maps of the nucleus of NGC1068.
The K$_n$ data are shown at the top, H in the middle and J at the bottom.
A vector 1 arcsecond long represents 5\% polarisation.  The contours
are of the total flux and are scaled arbitrarily.  

\clearpage

\begin{center}\begin{minipage}{5in}
\psfig{file=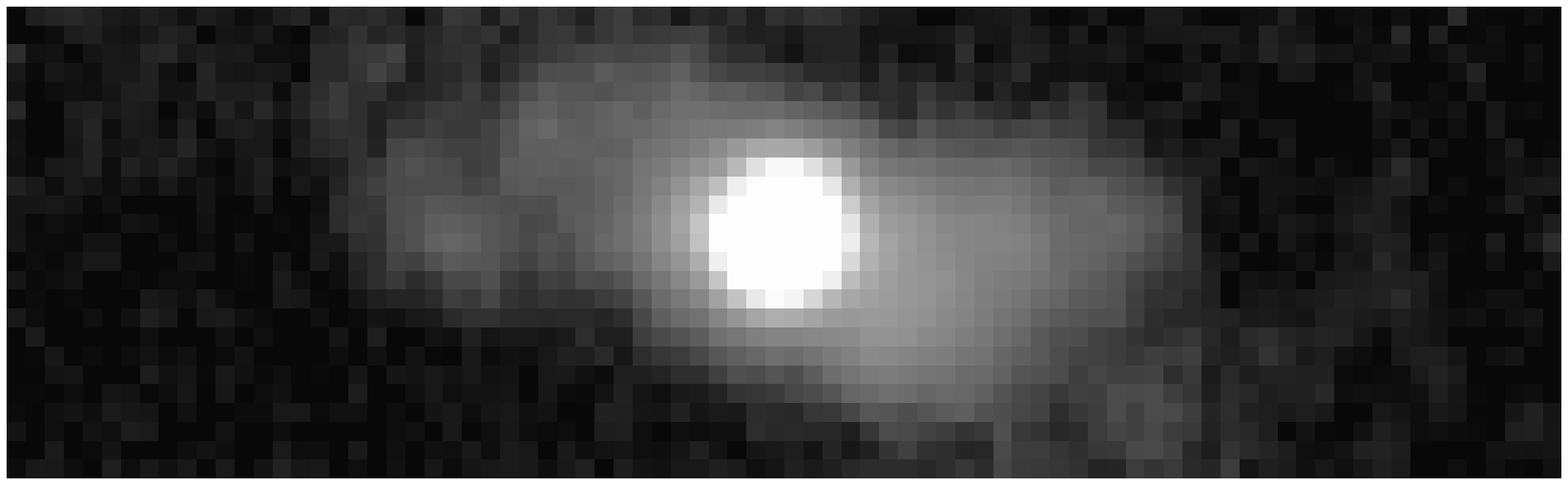,angle=0,width=5in,bblly=400pt,bbury=570pt,bbllx=15pt,bburx=550pt,clip=}

\vspace*{-1.4in}
\hspace*{0.2in}\psfig{file=rotwh.eps,width=1.5cm}
\vspace*{1.1in}

\psfig{file=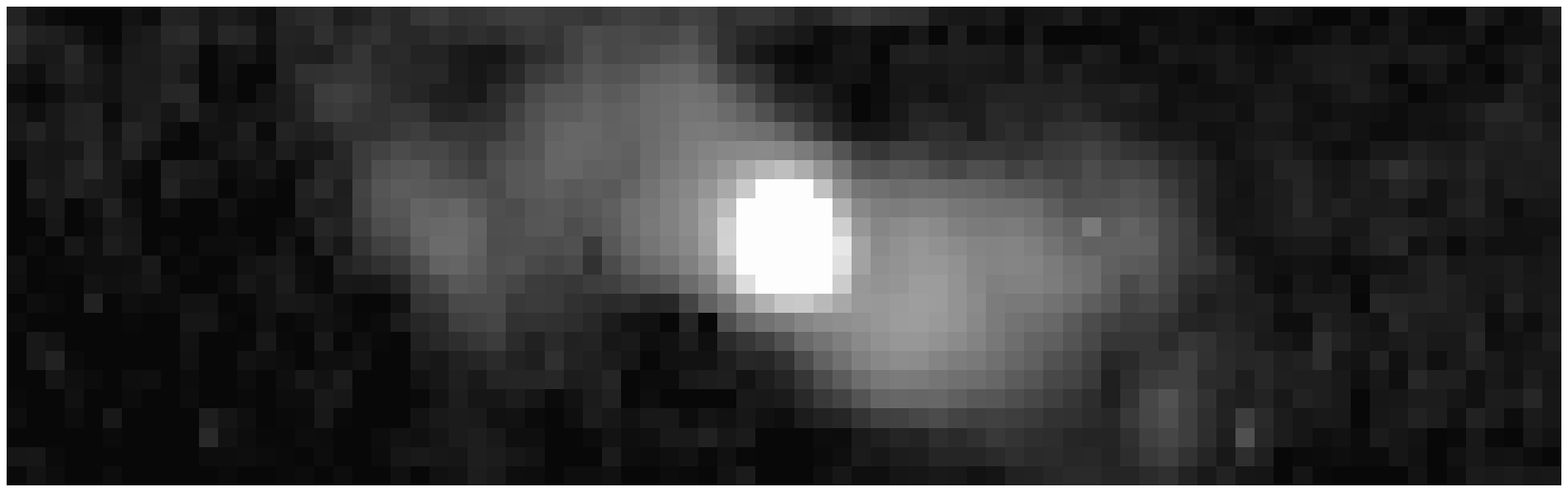,angle=0,width=5in,bblly=400pt,bbury=570pt,bbllx=15pt,bburx=550pt,clip=}

\psfig{file=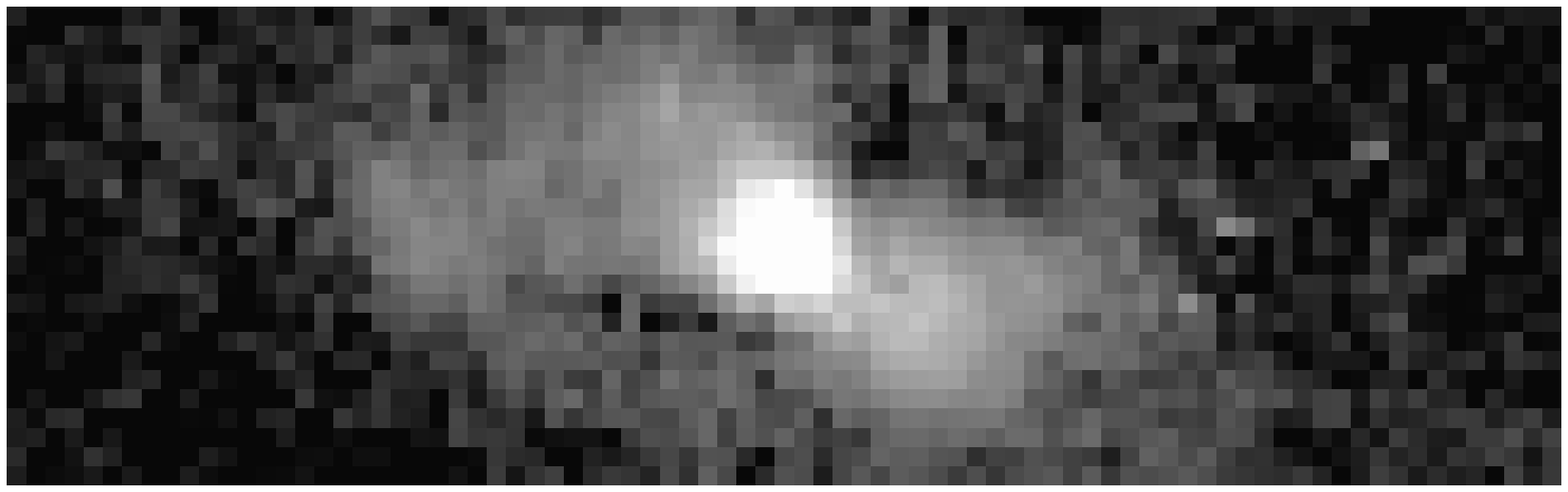,angle=0,width=5in,bblly=400pt,bbury=570pt,bbllx=15pt,bburx=550pt,clip=}
\end{minipage}\end{center}

{\bf Figure 2:} Near infrared polarisation images of NGC1068.
The K$_n$ data are shown at the top, H in the middle and J at the bottom.
The scale is the same as in Figure 1.

\vfill\pagebreak

\begin{center}\begin{minipage}{5in}
\psfig{file=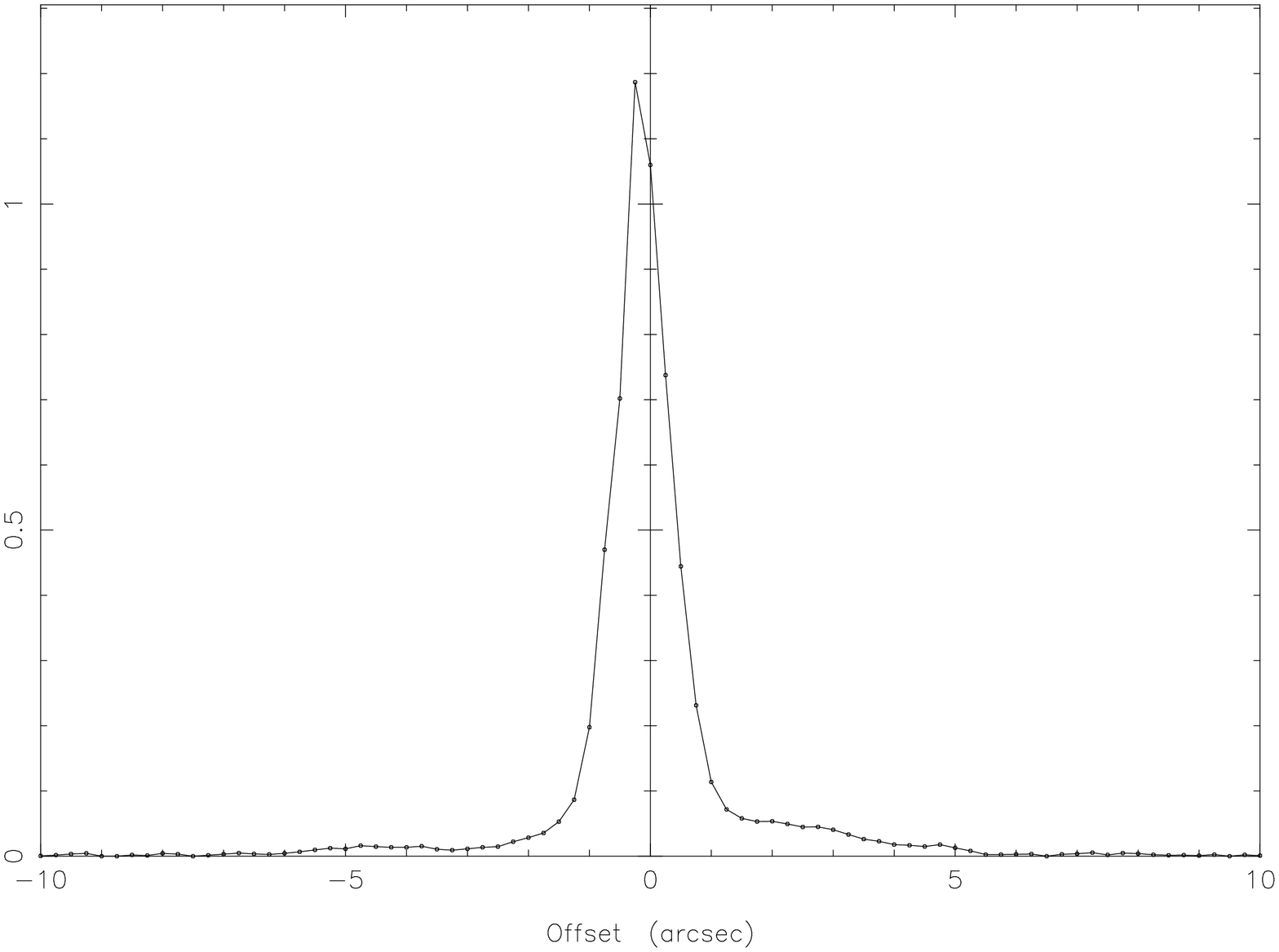,angle=-90,width=4in}
\vspace*{-.2in}
\psfig{file=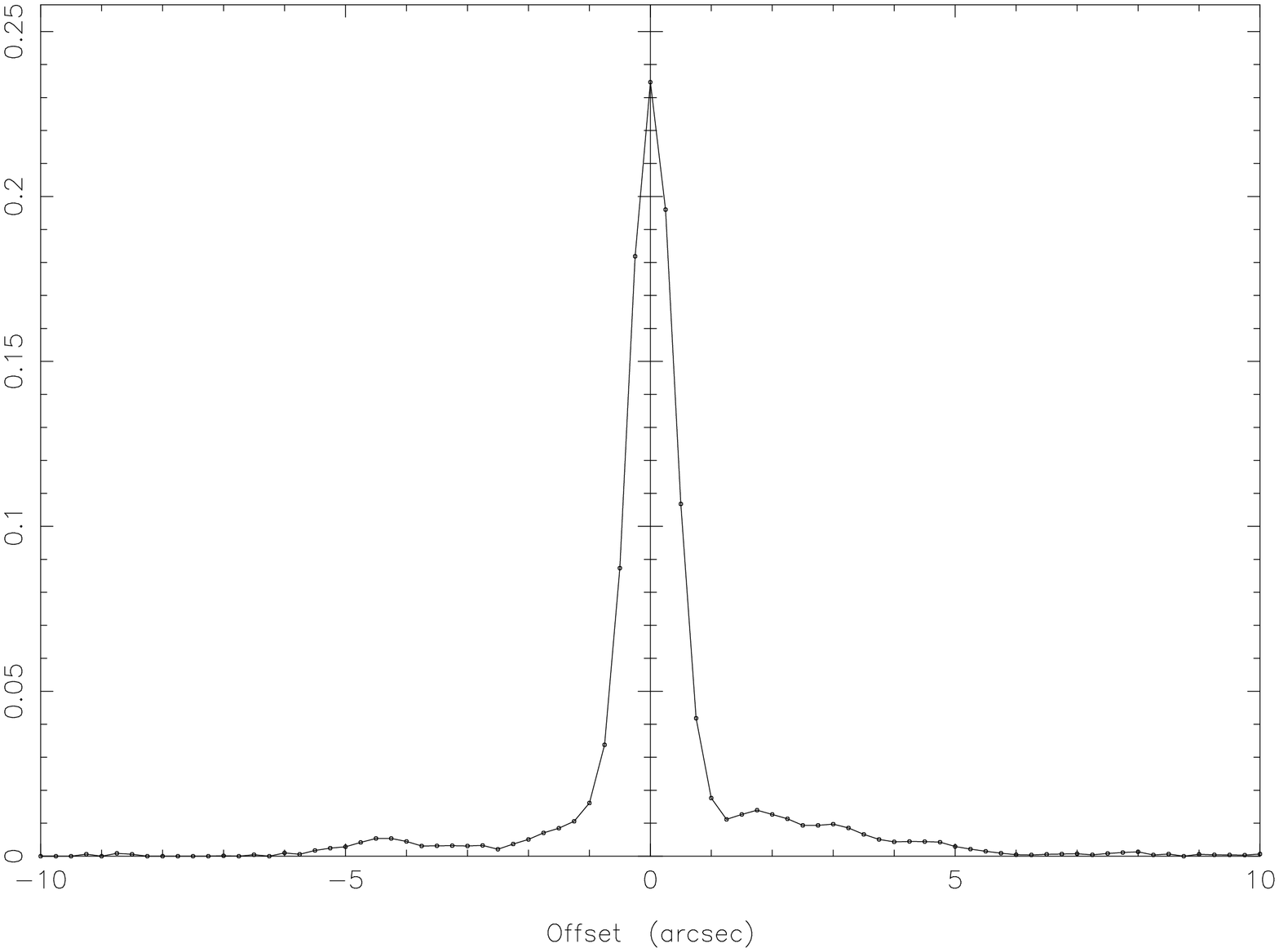,angle=-90,width=4in}
\vspace*{-.2in}
\psfig{file=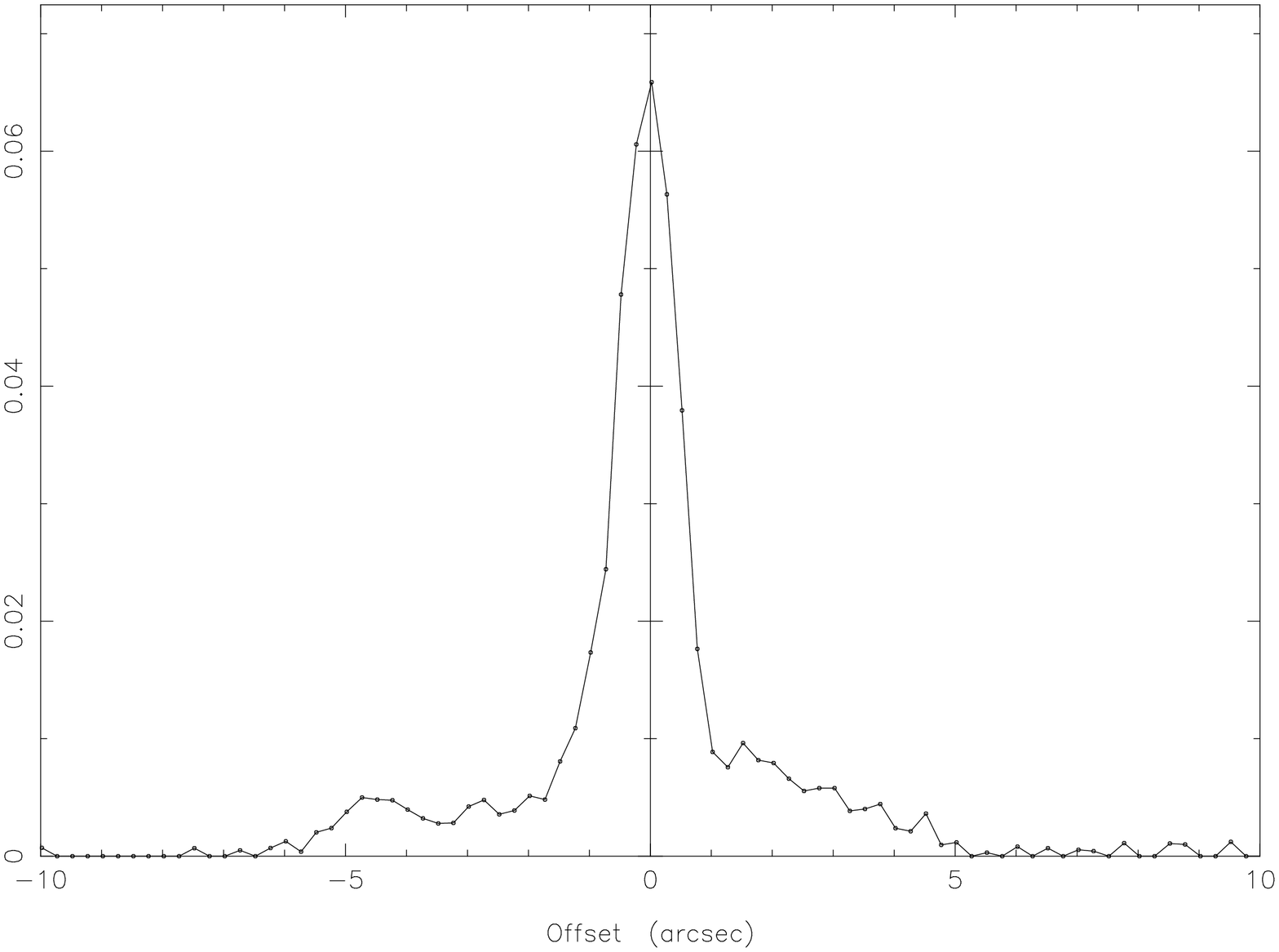,angle=-90,width=4in}
\end{minipage}\end{center}
\vspace*{.2in} {\bf Figure 3:} Cross section through the images shown in
Figure 2.  The K$_n$ data are shown at the top, H in the middle and J at the
bottom.  The dip in the profile at $-1$ arcsecond offset at H and J is the same
as the feature reported by Young et al.\ (1996)

\clearpage

\vspace*{7in}
{\bf Figure 4:} Mid infrared polarisation map of the nucleus of NGC1068.  The
data in the greyscale image have been smoothed slightly to enhance the contrast
in the faint extended emission.  The data are displayed on a power-law scale
(surface brightness to the third power).  The contours are evenly distributed
in on the same scale, but are otherwise arbitrary.
\vspace*{-8in}

\begin{center}\begin{minipage}{5in}
\psfig{file=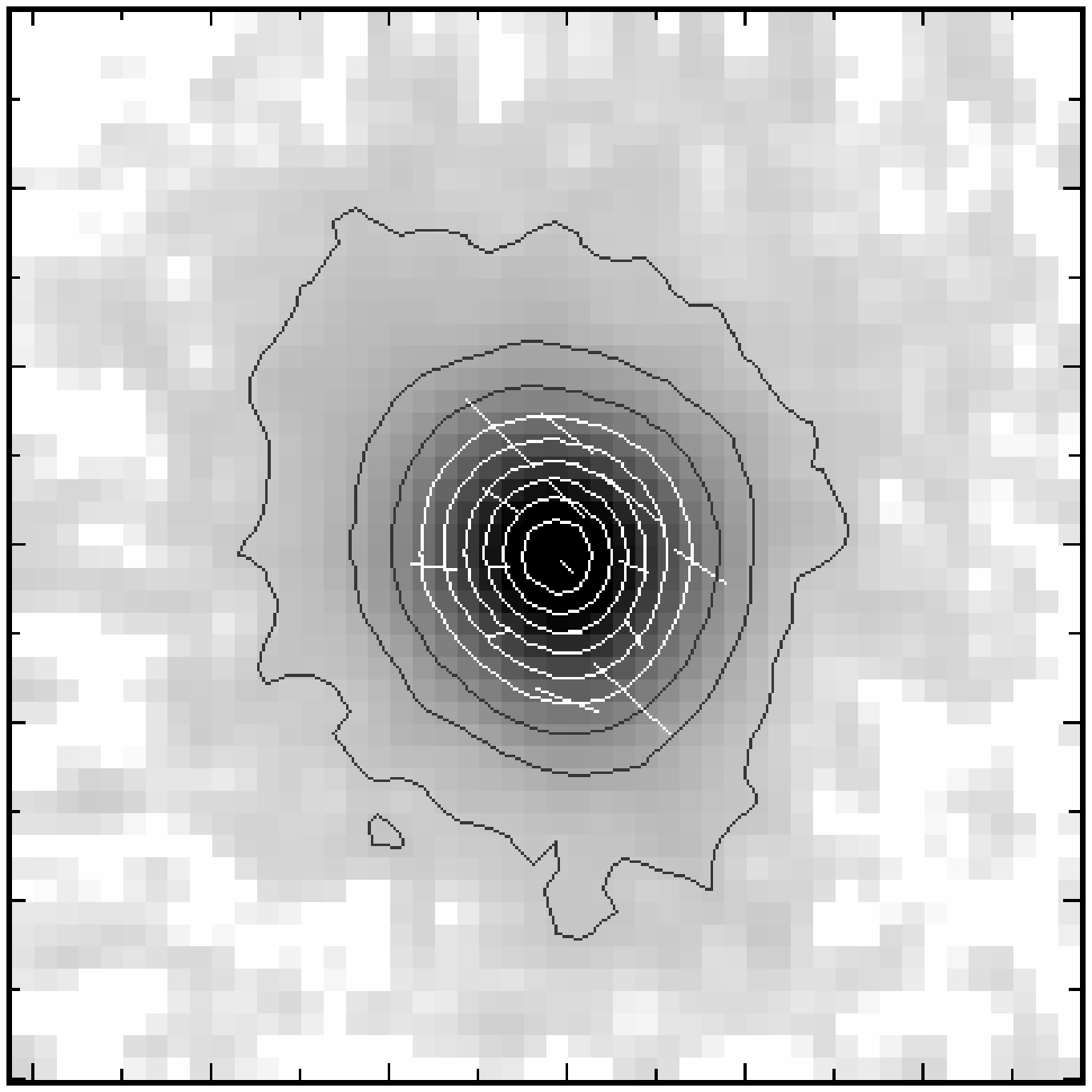,angle=0,width=5.24in,bbllx=97pt,bburx=502pt,bblly=225pt,bbury=630pt,clip=}
\vspace*{-4.4in}\hspace*{-12mm}
\psfig{file=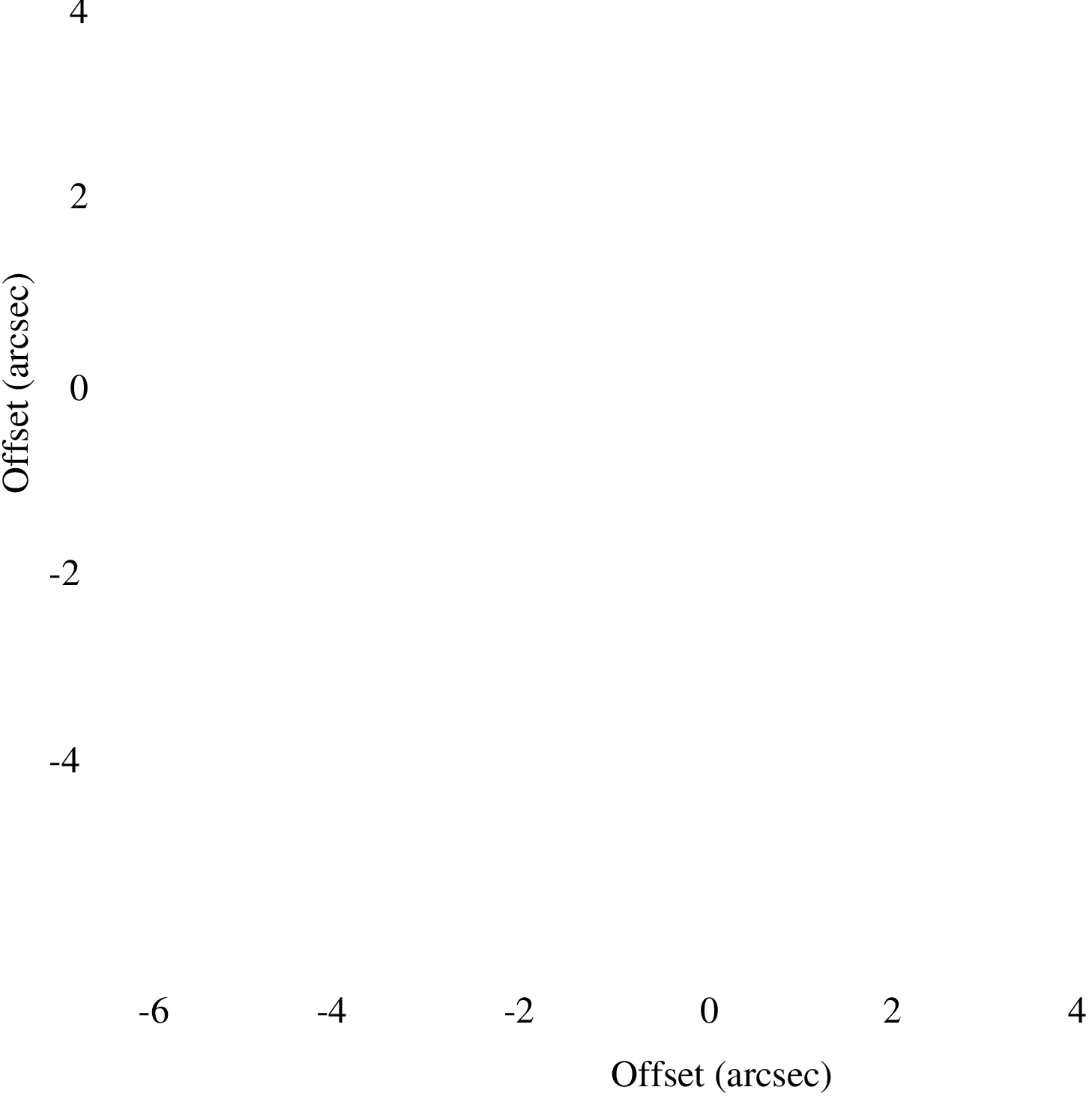,angle=-90,width=4.9in}
\end{minipage}\end{center}

\vspace*{-6.5in}
\hspace*{1.3in}\psfig{file=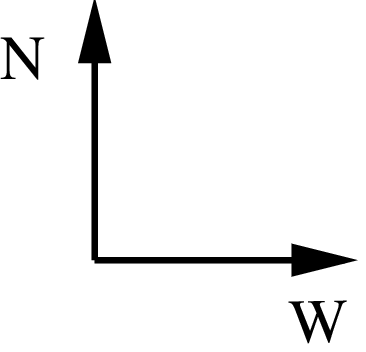,angle=-90,width=1.5cm}
\vspace*{6.9in}

\phantom{ggggg}

\begin{center}\begin{minipage}{5in}
\psfig{file=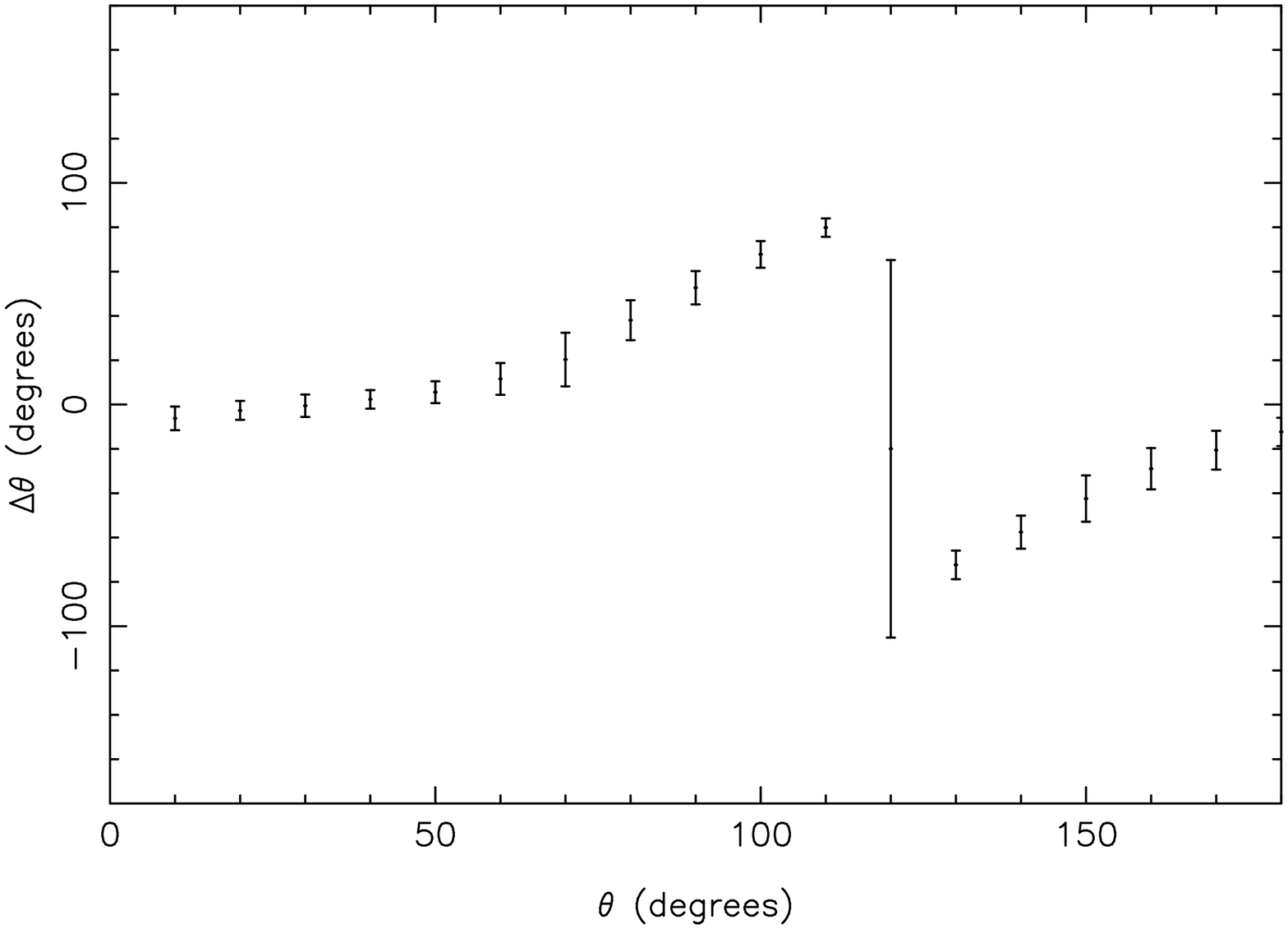,angle=-90,width=4in}
\vspace*{-.2in}
\psfig{file=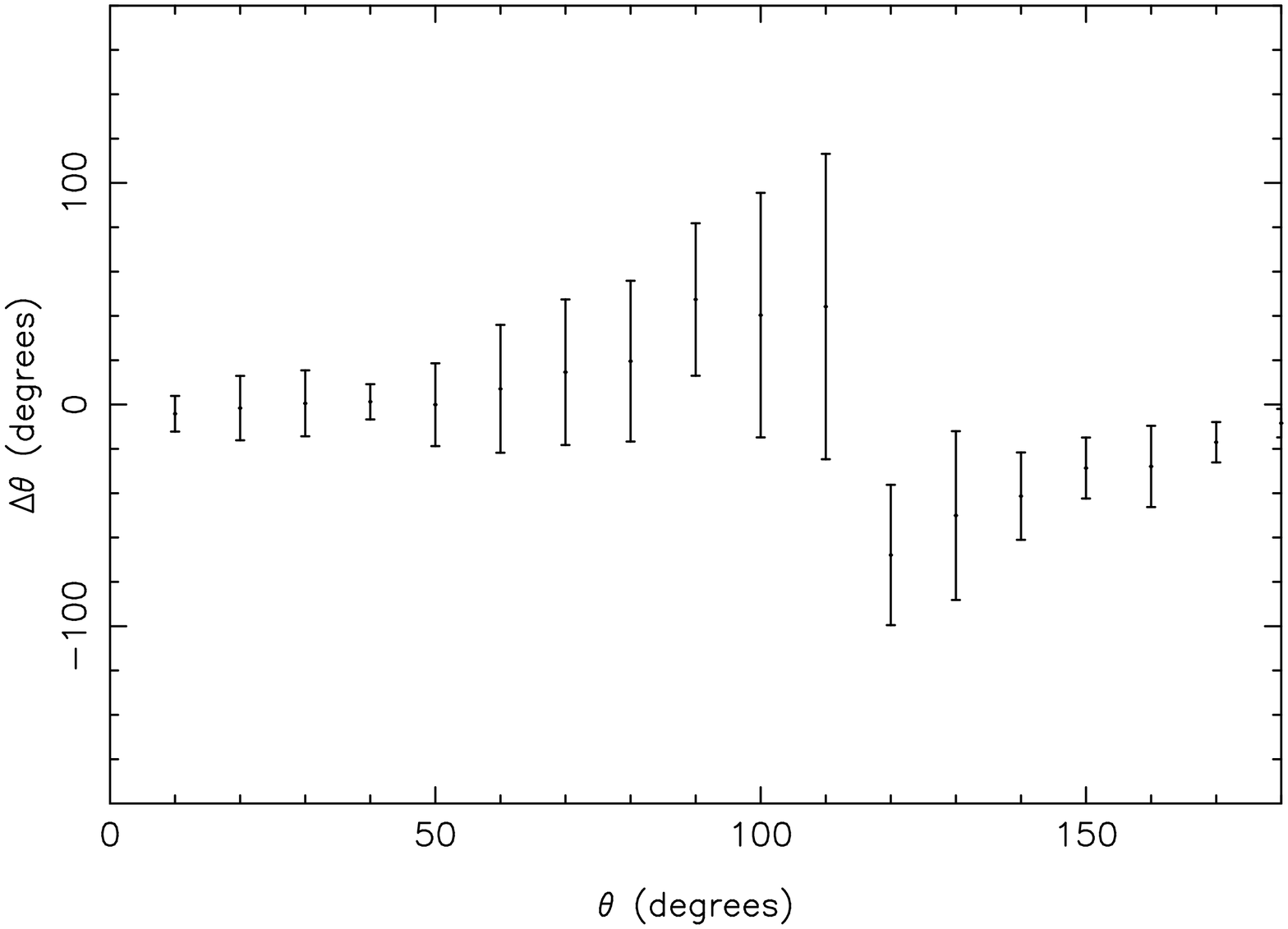,angle=-90,width=4in}
\vspace*{-.2in}
\psfig{file=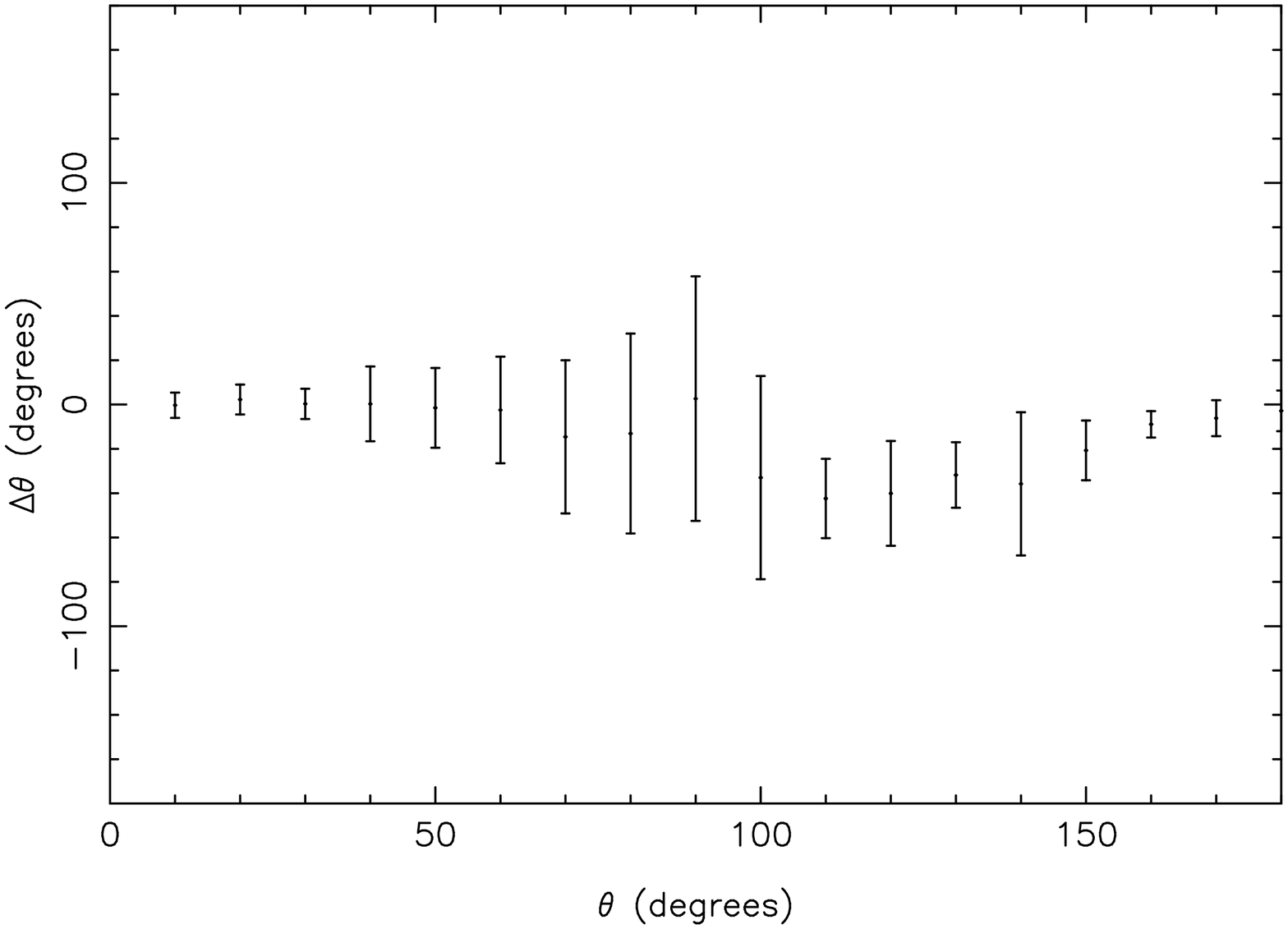,angle=-90,width=4in}
\end{minipage}\end{center}
{\bf Figure 5:} Deviation of the observed position angle of polarisation from
that expected from a centro-symmetric scattering pattern.  Pure scattering
would be represented by a straight line through zero.  
Again, the K$_n$ data are shown at the top, H in the middle and J at the
bottom.
The greatest
departure from zero is always at approximately 115$^\circ$.  The width
of this feature in the J band data is a reflection of the relative weakness
of the additional polarisation component in this case.

\end{document}